\documentclass[twocolumn,times]{aastex63}

\usepackage{newtxtext,newtxmath}
\usepackage{graphicx}	
\usepackage{amsmath}	
\usepackage{amssymb}	
\usepackage{mathrsfs}

\submitjournal{ApJL}

\usepackage{color}

\definecolor{blazeorange}{rgb}{1.0, 0.4, 0.0}
\definecolor{seagreen}{rgb}{0.18, 0.55, 0.34}
\definecolor{rufous}{rgb}{0.66, 0.11, 0.03}
\definecolor{royalfuchsia}{rgb}{0.79, 0.17, 0.57}
\definecolor{scarlet}{rgb}{1.0, 0.13, 0.0}
\definecolor{royalpurple}{rgb}{0.47, 0.32, 0.66}

\def\beq{\begin{equation}}
\def\eeq{\end{equation}}
\def\beqn{\begin{eqnarray}}
\def\eeqn{\end{eqnarray}}

\shorttitle{Probing BNS Merger EOS with FRBs}
\shortauthors{Yamasaki, Totani \& Kiuchi}

\begin{document}

\title{FRB 181112 as a Rapidly-Rotating Massive Neutron Star just after a Binary Neutron Star Merger?:\\ Implications for Future Constraints on Neutron Star Equations of State}

\correspondingauthor{Shotaro Yamasaki}
\email{shotaro.yamasaki@mail.huji.ac.il}
\author[0000-0002-1688-8708]{Shotaro Yamasaki}
\affiliation{Racah Institute of Physics, The Hebrew University of Jerusalem, Jerusalem 91904, Israel}
\affiliation{The Raymond and Beverly Sackler School of Physics and Astronomy, Tel Aviv University, Tel Aviv 69978, Israel}

\author{Tomonori Totani}
\affiliation{Department of Astronomy, School of Science, The University of Tokyo,  7-3-1 Hongo, Bunkyo-ku, Tokyo 113-0033, Japan}
\affiliation{Research Center for the Early Universe, School of Science, The University of Tokyo, 7-3-1 Hongo, Bunkyo-ku, Tokyo 113-0033, Japan}

\author{Kenta Kiuchi}
\affiliation{Max-Planck-Institut für Gravitationsphysik, Albert-Einstein-Institut, Am Mühlenberg 1, Potsdam-Golm D-14476, Germany}
\affiliation{Center for Gravitational Physics, Yukawa Institute for Theoretical Physics, Kyoto University, Kyoto 606-8502, Japan}



\begin{abstract}

The light curve of the fast radio burst (FRB) 181112 is resolved into four successive pulses, and the time interval ($\sim0.8$ ms) between the first and third pulses coincides with that between the second and fourth pulses, which can be interpreted as a neutron star (NS) spinning at a period of about $0.8$ ms. 
Although this period is shorter than the most rapidly rotating pulsar currently known (1.4 ms), it is typical for a simulated massive NS formed immediately after the coalescence of binary neutron stars  (BNS).
Therefore, a BNS merger is a good candidate for the origin of this FRB if the periodicity is real. We discuss the future implications that can be obtained if such a periodicity is detected from FRBs simultaneously with gravitational waves (GW). The remnant spin period $P_{\rm rem}$ inferred from the FRB observation is unique information which is not readily obtained by current GW observations at the post-merger phase. If combined with the mass of the merger remnant $M_{\rm rem}$ inferred from GW data, it would set a new constraint on the equation of state of nuclear matter. Furthermore, the post-merger quantity $P_{\rm rem}/M_{\rm rem}$, or the tidal deformability of the merger remnant, is closely related to the binary tidal deformability parameter $\Tilde{\Lambda}$ of NSs before they merge, and a joint FRB-GW observation will establish a new limit on $\Tilde{\Lambda}$. Thus, if $\Tilde{\Lambda}$ is also well measured by GW data, a comparison between these two will provide further insights into the nature of nuclear matter and BNS mergers.

\end{abstract}


\section{Introduction}

Fast radio bursts (FRBs) are cosmological radio transients with millisecond duration whose origin is enigmatic \citep{Lorimer2007,Thornton2013}.
In the last decade there has been a remarkable increase in our knowledge
about FRBs from the observational perspective \citep{Petroff2019,Cordes2019}. Identification of the host galaxy with sub-arcsecond localization accuracy for a dozen of FRBs has placed these sources at redshifts between $0.03$ and $0.66$ \citep{Chatterjee2017,Bannister2019,Ravi2019b,Prochaska2019,Marcote2020,Bhandari2020,Heintz2020}, confirming their cosmological origin. While it has been argued that all FRB sources could potentially repeat \citep[e.g.,][]{Ravi2019a}, 
some sources, such as FRB 121102 \citep{Spitler2016}, are statistically more active than the others \citep[e.g.,][]{Law2017,Palaniswamy2018} and intrinsic burst widths for the repeating sources are known to be larger as compared to those for the so-far non-repeating sources \citep{CHIME2019,Fonseca2020}, supporting the
notion of potentially different origins the two.

The spectral and polarimetric properties of FRBs at high time resolution are crucial to understanding their emission mechanisms and local environments \citep[e.g.,][]{Farah2018,Hessels2019,Nimmo2020}. One of such sources, thus-far non-repeating FRB 181112, was detected in the Commensal Real-time ASKAP Fast Transients (CRAFT) survey at $1.3$ GHz, with a duration of $2.1$ ms and a fluence of $26\ {\rm Jy\,ms}$, as reported by \citet{Prochaska2019}. The burst was localized to a star-forming galaxy at redshift $z=0.48$. Nevertheless, it has a linear polarization at negligible Faraday rotation measure (RM) $\sim10\ {\rm rad\  m^{-2}}$, which may disfavor an extremely magneto-ionic environment, such as a supernova remnant.
Recently, \citet{Cho2020} carried out a high time ($3$-ns) resolution analysis of this burst, and found out that the burst is composed of a train of four narrow pulses separated by submilliseconds each. More recently, \citet{Day2020} also found that double-peaked FRBs 190102 and 190611 detected by ASKAP, both share some phenomenological similarities with FRB 181112.

In this {\it Letter}, we consider implications of the observations of narrow pulses in FRB 181112. Intriguingly, the time interval between the first and third pulses ($\sim0.8$ ms) coincides with that between the second and fourth pulses, implying that there would be an underlying neutron star (NS) spinning at extremely short period $\sim0.8$ ms, and such a fast rotation could be most naturally achieved by a coalescence of binary neutron stars (BNS). In this case, spinning magnetic fields of merging NSs \citep{Totani2013} or the interaction between the NS magnetospheres \citep{Wang2016} during the final stage of a BNS merger inspiral can produce an FRB, which would be hidden at $0.5$-$1$ ms after the merger due to the subsequent mass ejection \citep{Yamasaki2018}.
Therefore, if our interpretation of the underlying periodicity between sub-pulses of FRB 181112 is correct, it would provide strong support to the BNS-merger origin for this FRB. 
We also examine the idea that future co-detection of the gravitational wave (GW) from such FRBs provide the completely new information on the NS that is complementary to the  relatively poor sensitivity of current 
GW  detectors right after the BNS merger. 

This {\it Letter} is organized as follows. In \S \ref{s:FRB181112}, we present our interpretation of the temporal properties of FRB 181112 on the basis of BNS merger model for FRBs. In \S \ref{s:constraints}, possible constraints on  the NS equations of state is presented, followed by discussion in \S  \ref{s:discussion}. Throughout this work, we use a geometrical unit with $c=G=1$, where $c$ and $G$ are the speed of light and the gravitational constant, respectively.

\section{FRB 181112 from BNS merger?}
\label{s:FRB181112}

\subsection{Interpretation of Four Narrow Pulses}
\label{ss:period}
According to \cite{Cho2020}, the singly-detected FRB 181112 consists of four narrow pulses with signal-to-noise ratio (S/N) of $220$, $5$, $28$, and $8$ respectively, arriving at times $t_1=0$ ms, $t_2=0.48\pm0.01$ ms, $t_3=0.808\pm0.004$ ms, and $t_4=1.212\pm0.002$ ms, where $t_i$ refers to the peak of the profile of pulse $i$. 
Although no significant periodicity cannot be claimed because of the small number of pulses,
the interval $t_{31}$ = 0.808 ms between the pulses 1 and 3
is interestingly close to $t_{42}$ = 0.732 ms between the pulses 2 and 4,
implying a tempting possibility of a periodicity around 0.8 ms \citep{Cho2020}. This is consistent with the duration (width) of an individual pulse $\lesssim0.1$ ms and negligible temporal pulse broadening due to scattering $\sim20\,\mu$s. 
Assuming that the four pulses randomly distribute within the time window of $2P\sim1.6$ ms, we estimate by a simple Monte Carlo method that there is a non-negligible $\lesssim18$\% probability for 
the $|t_{31}-t_{42}|$ to be less than the observed difference $0.076$ ms by chance alone. 

Though this chance probability alone does not allow us to claim the existence of periodicity with a high level of confidence,
there is other circumstantial evidence to support the hypothesis that pulses 1 and 3 are of the same origin (see, e.g., Table 1 and Figure 1 of \citealt{Cho2020}). First, pulses 1 and 3 have similarly high S/N, which is in contrast to the low S/N of weak pulses 2 and 4. Secondly, the Faraday rotations were measured only in pulses 1 and 3 with their polarization angles being consistent with each other within $20^{\circ}$, suggesting that a similar magnetic field geometry may have been achieved in these two pulses. 
Furthermore, there is a potential similarity in the time-frequency structures of pulses. The dynamic spectra of pulses 1 and 3 extend across the observing band up to $\sim1450$ MHz, whereas pulses 2 and 4 possibly have a spectral cutoff at around $\lesssim1300$ MHz (see Figure 2 of \citealt{Cho2020}).
Therefore, these data might support the interpretation of the $0.8$ ms periodicity, in which case the four pulses are emitted over two rotational periods.

If the submillisecond periodicity is real, it is reasonable to regard it as the rotation of an underlying compact star, such as a NS.
\citet{Lattimer2007} showed that the minimum spin period for a uniformly rotating NS
with non-rotating mass $M$ and radius $R$ in fully relativistic calculations employing realistic hadronic EOSs is approximated as $P_{\rm min}\approx(0.96\pm0.03)\left(M_{\odot}/M\right)^{1/2}\left(R/10\ {\rm km}\right)^{3/2}$ ms, which applies to an arbitrary NS mass as long as it is not close to the maximum non-rotating mass (whereas for the maximum mass configuration, the coefficient reduces to $0.83$). 
That is, for $M\gtrsim1.4\ M_{\odot}$ star with its radius $R=10$ km, the minimum spin period would be limited to $P_{\rm min}\lesssim0.8$ ms regardless of EOS.   
Thus, most NS EOSs, at least in theory, allow for short spin period at $P=0.8$ ms seen in FRB 181112.

\subsection{The Origin of Most Rapidly Spinning NS}

What is the possible progenitor of a NS with such a short spin period? The most common channel for the formation of NSs is core collapse supernovae (CCSNe).
Current evolutionary models of progenitors combined with numerical simulations of core collapses and explosions \citep[e.g.,][]{Spruit1998,Heger2000,Heger2005,Thompson2005,Ott2006,Nakamura2014} show that the spin period of a newborn NS can be as small as a few milliseconds only if the spin rate of the progenitor is sufficiently high.
Meanwhile, the initial spins of pulsars are not well constrained by observations but, most likely, they lie in the vicinity of tens to hundreds of milliseconds
\citep[e.g.,][]{Narayan1987,Lorimer1993,Kaspi2002,Faucher2006,Miller2015}.
The most rapidly rotating pulsar currently known is J1748--2446ad with $P=1.4$ ms \citep{Hessels2006}, which however is not an isolated pulsar but a recycled one in a binary system, and so far no submillisecond pulsars have been found, despite vigorous pulsar explorations \citep[e.g.,][]{Lorimer2008}.

Moreover, depending on the mass of SN ejecta, it takes about $10$--$100$ years for the surrounding environment to become transparent to the radio waves \citep[e.g.,][]{Murase2016,Kashiyama2017,Metzger2017}. Thus, even though the remnant NS from a CCSN is born rapidly rotating, with an initial spin period of submilliseconds, it would have significantly spun down by the time when an FRB produced by its activity could be observed, unless the initial NS magnetic field is too low and/or the NS angular momentum significantly increases due to a fallback accretion   \citep[e.g.,][]{Shigeyama2018}.
Therefore, the explanation of the submillisecond rotation by CCSNe requires fine tuning of the parameters.

Another possible channel for NS formation is the coalescence of binary neutron stars (BNSs). Such a remnant after the merger, called massive NS, would start out rapidly rotating and gradually slow down through emission of gravitational and electromagnetic radiations \citep{Shibata2019}
mass and also on the NS equations of
state (EOS), it could survive for hundreds of milliseconds and eventually collapse to a black hole \citep{Hotokezaka2013} or it could actually remain stable indefinitely \citep{Shibata2019}.
Since the remnant NS inherits the large kinetic energy $\sim10^{53}\ {\rm erg}$ of the binary orbital motion, its initial spin period is typically about $0.5$--$1.0$ ms, which is suggested by the numerical relativity simulations with the plausible value of the binary mass of $2.5$--$2.7M_\odot$ \citep{Radice2018}. In this respect, the explanation of the $P=0.8$ ms rotation seen in FRB 181112 would be most naturally interpreted as the spin rate of the BNS merger remnant without fine tuning of parameters.

Furthermore, in the framework of BNS merger scenarios for FRBs \citep{Totani2013,Wang2016}, the rotational energy budget available for FRB emission dramatically increases until the moment of coalescence. Meanwhile, the dynamical ejecta begin to screen the radio emission at times about $0.5$--$1$ ms after the merger \citep{Yamasaki2018}, which may limit the maximum duration of an FRB and thus one will not ``see'' the subsequent FRB sub-pulses if any\footnote{If the remnant NS survives for long time ($\gtrsim1$ year) after the merger, its rotational or magnetic activity may produce repeating FRBs (\citealt{Yamasaki2018}, see also \citealt{Margalit2019,WangFY2020}).}. Therefore, we conclude that FRB 181112 could be most naturally interpreted as the repeated radio emissions from the remnant NS around the moment of coalescence that have survived the absorption due to the subsequent expansion of dynamical ejecta.

\section{Future Implications on Neutron Star Equations of State}
\label{s:constraints}

While the possible presence of submillisecond periodicity in FRB 181112 strengthens the support for the BNS merger origin for this FRBs as shown in \S \ref{s:FRB181112}, the most unambiguous confirmation is only achieved by detecting the GW emission simultaneously with an FRB \citep{Totani2013,Zhang2014,Yamasaki2018,WangMH2020}.
In this section, we discuss the future implications of a simultaneous detection of an FRB 181112-like FRB and the associated GW for NS matter EOSs. For these purposes we make use of the latest numerical-relativity simulations of BNS mergers (\S \ref{ss:simulation}). Based on this, we demonstrate some relations among key BNS-merger properties and show how FRB and GW observations can be combined with such relations to constrain the NS properties (\S \ref{ss:relations} and \S \ref{ss:Lambda}).

\subsection{Simulation Data and Physical Quantities of Interest}
\label{ss:simulation}

We use the numerical-relativity simulations of BNS mergers \citep{Kiuchi2017,Kiuchi2020} performed with five phenomenological EOSs
(polytropic EOSs for dense nuclear matter with broken power law, see \citealt{Read2009}), which produce a wide range of spherical NS radii $R_{1.35}=10.96$--$13.69$ km for a $1.35\ M_{\odot}$ star. As shown below, our purposes are to demonstrate the qualitative dependence of the remnant spin period on the remnant mass and to obtain the relationship between compactnesses before and after the merger. Therefore, a choice of relatively simple EOSs is sufficient.
Here we try the models with total mass $m_{\rm tot}=2.50$--$2.73\,M_{\odot}$ and mass ratio $q=m_1/m_2=0.7$--$1.0$, where $m_1+m_2=m_{\rm tot}$. The primary quantities of our interest are the minimum spin period of the remnant ($P_{\rm rem}$), the remnant mass ($M_{\rm rem}$), and the binary tidal deformability parameter ($\tilde{\Lambda}$), which are extracted from the simulations as follows. 

In general, the remnant is initially rotating differentially, which is characterized by a slowly rotating core surrounded by a rapidly rotating outer layer, and depending on the magnetorotational instabilities and/or the neutrino cooling , the rotational profile evolves into Keplerian one \citep[e.g.,][]{Shibata2005,Fujibayashi2020}.
Namely, the rotational profile is highly unstable (hence not appreciable) around the time of merger.
Thus, we extract the minimum spin period by examining the location of the peak in the angular velocity profile along the equator of the merged NS remnant (or orbital plane) at about $10$--$15$ ms after the merger.  The errors in $P_{\rm rem}$ arising from simulations are estimated to be $\lesssim6$\%. We approximate the remnant mass by the total mass of the NSs for simplicity (i.e., $M_{\rm rem}\sim m_{\rm tot}$). Other potential systematic uncertainties in $P_{\rm rem}$ and $M_{\rm rem}$ will be discussed in \S \ref{s:discussion}. 
This is reasonable because the total mass of tidal and shock-driven dynamical ejecta during the early post-merger phase is typically  $\lesssim10^{-2}\ M_{\odot}$ \citep[e.g.,][]{Shibata2019}, which is negligible compared to the total mass of the system. 
Last but not least, the binary tidal deformability $\tilde{\Lambda}$ is directly extracted from the inspiral GWs \citep[][]{LIGO2017,De2018,De2018Erratum,LIGO2020}.

\begin{figure}[t]
\centering
\includegraphics[width=.48\textwidth]{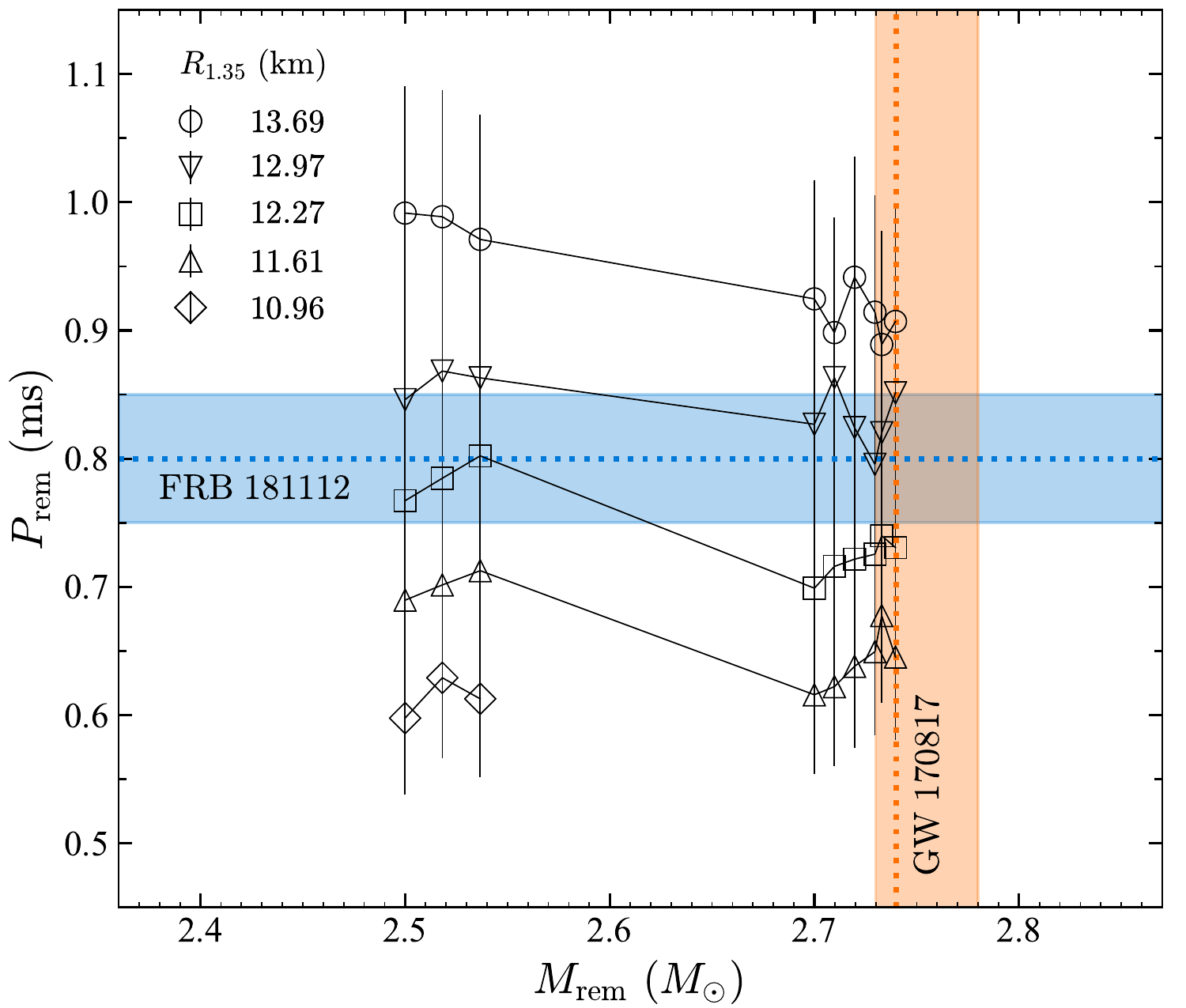}
\caption{The post-merger spin period of the remnant NS $P_{\rm rem}$ as a function of remnant mass $M_{\rm rem}$, which is approximated as a total mass of the NSs. The vertical error bars are coming from the simulation uncertainties arising from the differential rotation of the remnant, which are set to be $6$\%. 
The different markers represent different EOSs and the non-rotating spherical NS radii for a $1.35\,M_\odot$ could be regarded as an effective parameter of the EOS. The simulation with softest EOS (corresponding to $R_{1.35}=10.96$ km) with high total mass $m_{\rm tot}\gtrsim2.7M_{\odot}$ are not shown because they collapse to a black hole within a few ms after the merger and hence $P_{\rm rem}$ is unavailable. The horizontal blue region represents the potential spin period estimated from FRB 181112 observation with an error of $0.1$ ms (see \S \ref{ss:period}) and the vertical orange region indicates the total mass of the BNS system inferred from the GW170817 observations $2.74_{-0.01}^{+0.04}\,M_{\odot}$ \citep{LIGO2017}.
}
\label{figure_1}
\end{figure}

\begin{figure*}[t]
\centering
\includegraphics[width=.75\textwidth]{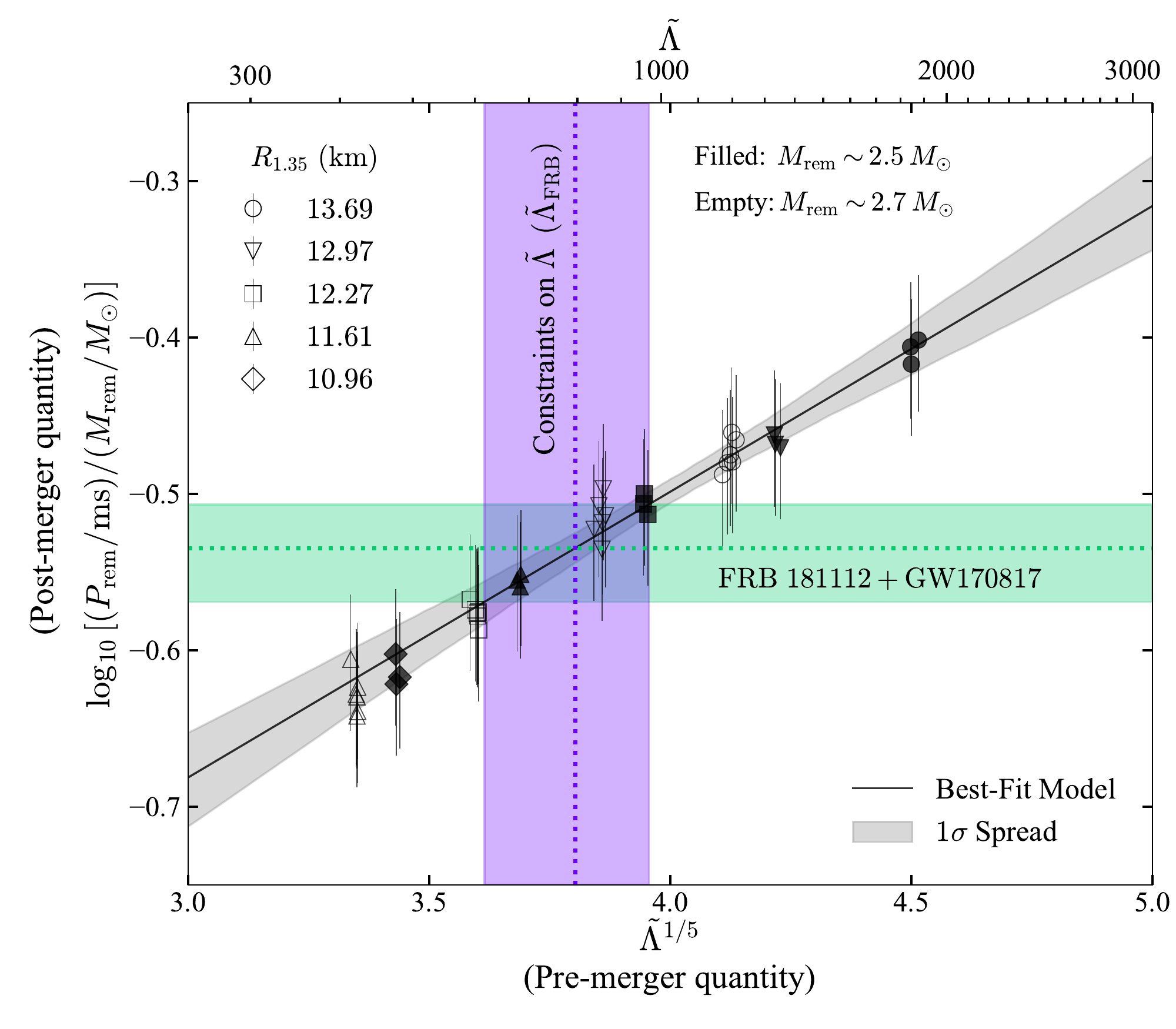}
\caption{$P_{\rm rem}/M_{\rm rem}$ as a function of binary tidal deformability parameter ${\Tilde{\Lambda}}^{1/5}$ (corresponding ${\Tilde{\Lambda}}$ values are indicated above the upper horizontal axis). The best-fitted result is shown as a
solid black line with grey shaded region being the 1-$\sigma$ errors. The  horizontal green region represents the $P_{\rm rem}/M_{\rm rem}$ value estimated from the hypothetical detection of an FRB-GW event (i.e., FRB 181112 and GW170817). The vertical purple region represents the constraints on the tidal deformability coming from the $P_{\rm rem}/M_{\rm rem}$ constraints. 
}
\label{figure_2}
\end{figure*}

\subsection{Period-Mass Relation}
\label{ss:relations}

Figure \ref{figure_1} shows the relation between $P_{\rm rem}$ and $M_{\rm rem}$ 
for different EOSs. One can see that the $P_{\rm rem}$-$M_{\rm rem}$ relation strongly depends on the EOS (or radius $R_{\rm rem}$), and for each EOS there is a mild dependence of $P_{\rm rem}$ on $M_{\rm rem}$. These trends could be qualitatively understood if the remnant has a quasi-uniform rotation and it rotates with the Keplerian velocity at the surface, i.e., $P_{\rm rem}\propto R_{\rm rem}^{3/2}\,M_{\rm rem}^{-1/2}$.
This plot is useful when considering a case of detecting the inspiral GW and coincidentally seeing a high-time-resolved FRB with submillisecond periodicity. In this case, $P_{\rm rem}$ and $R_{\rm rem}$ are both measurable by the FRB and GW observations, respectively. For instance, let us consider a hypothetical FRB-GW detection by taking $P_{\rm rem}$ from FRB 181112 and $M_{\rm rem}$ from GW170817. Then, one can constrain the allowed parameter space on the $P_{\rm rem}$-$M_{\rm rem}$ plane (see the area where the horizontal and vertical shaded regions intersect in Figure \ref{figure_1}).
This demonstrates that the simultaneous measurement of $P_{\rm rem}$ and $M_{\rm rem}$ would provide an important constraint on the EOS.

\subsection{Period/Mass-Tidal Deformability Relation}
\label{ss:Lambda}

Additionally, we consider a case where the tidal deformability (as well as $M_{\rm rem}$)
is measured by the GW observations observation of a BNS inspiral.
The binary tidal deformability, $\Tilde{\Lambda}$, can be written as \citep[][]{Flanagan2008,Hinderer2007}
\beqn
\label{eq:Lambda}
\Tilde{\Lambda}&=&\frac{8}{13}[(1+7\eta-31\eta^2)(\Lambda_1+\Lambda_2)\nonumber\\ 
&+&\sqrt{1-4\eta}(1+9\eta-11\eta^2)(\Lambda_1-\Lambda_2)],
\eeqn
where $\eta=m_1m_2/m_{\rm tot}^2$ is the symmetric mass ratio and $\Lambda_i$ ($i=1,\,2$) is the tidal deformability of each star, defined as
\beq
\label{eq:Lambda_i}
\Lambda_i\equiv\frac{2}{3}k_2^{(i)}\left(\frac{R_i}{m_i}\right)^5,
\eeq
where $k_2^{(i)}$ is the quadrupolar Love numbers of each NS. For simplicity, we consider near-equal-mass NSs with $\eta\sim1/4$, in which case $\Tilde{\Lambda}\sim\Lambda_i$. As shown in Figure \ref{figure_2}, one can see that the remnant quantity $P_{\rm rem}/M_{\rm rem}$ is closely related to the binary tidal deformability by the relation\footnote{This is qualitatively similar to the so-called ``(approximate) universal relations'' between the post-merger
gravitational wave frequency and tidal deformability \citep[e.g.][]{Bauswein2012,Read2013,Bernuzzi2015,Rezzolla2016,Zappa2018,Kiuchi2020}}.
\beq
\label{eq:P/M-Lambda}
\log_{10}\left[\left(\frac{P_{\rm rem}}{\rm ms}\right)\left(\frac{M_{\odot}}{M_{\rm rem}}\right)\right]\simeq a_0 +a_1{\Tilde{\Lambda}}^{1/5},
\eeq
where $a_0=-1.22^{+0.08}_{-0.08}$ and $a_1=0.18_{-0.02}^{+0.02}$ are numerical coefficients with errors of 1-$\sigma$. This may be qualitatively understood as follows. By assuming that the remnant has a Keplerian rotation, $P_{\rm rem}/M_{\rm rem} \propto(R_{\rm rem}/M_{\rm rem})^{3/2}={\cal C}_{\rm post}^{-3/2}$, where ${\cal C}_{\rm post}$ is the compactness of the remnant NS. Meanwhile, the NSs' tidal deformability is related to the compactness of NSs before the merger ${\cal C}_{\rm pre}$ as $\tilde{\Lambda}^{1/5}\sim\Lambda_i^{1/5}\propto R_i/m_i={\cal C}_{\rm pre}^{-1}$ (see Eq. [\ref{eq:Lambda_i}])\footnote{A slightly different relationship between the binary tidal deformability and the compactness parameter $\tilde{\Lambda}\propto C^{-6}$ has also been proposed \citep{De2018,De2018Erratum}. Yet, this barely affects our conclusions.}.
Namely, each quantity could be expressed in terms of compactness parameter. 
Therefore, the clear $P/M$-$\tilde{\Lambda}^{1/5}$ correlation in Eq. \eqref{eq:P/M-Lambda} may imply the existence of hidden relationship between ${\cal C}_{\rm pre}$ and ${\cal C}_{\rm post}$, which could be only investigated through numerical-relativity simulations.

Given the remnant spin period and mass inferred by FRB and GW observations, respectively, one can see that Eq. \eqref{eq:P/M-Lambda} will provide a constraint on the tidal deformability $\Tilde{\Lambda}_{\rm FRB}$, which is completely independent from that directly measured from the inspiral GW $\Tilde{\Lambda}_{\rm GW}$. For instance with a hypothetical FRB-GW detection (taking $P_{\rm rem}$  
from FRB 181112 and $M_{\rm rem}$ from GW170817), one would obtain $\Tilde{\Lambda}_{\rm FRB}\sim600$--$1000$. This is actually consistent with the tidal deformability directly measured from the GW170817 $100\lesssim\Tilde{\Lambda}_{\rm GW}\lesssim800$ \citep{LIGO2017}. We note that the upper limit on the $\Tilde{\Lambda}_{\rm GW}$ is robustly set by the GW analysis, whereas the lower limit is rather dependent on the prior physical information about the NSs. In contrast, as our method of using $P/M$-$\tilde{\Lambda}^{1/5}$ relation provides a constraint on $\tilde{\Lambda}_{\rm FRB}$ (hence on NS radii) with an error bar, this would be qualitatively different estimate and thus of great importance. The error in $\tilde{\Lambda}_{\rm FRB}$ is subject to the accuracy of the FRB and GW observations ($P_{\rm rem}$ and $M_{\rm rem}$) and the variance of $P_{\rm rem}/M_{\rm rem}$-$\tilde{\Lambda}^{1/5}$ relation (see \S \ref{s:discussion}). 

Curiously, even the possible disagreement between $\Tilde{\Lambda}_{\rm FRB}$ and $\Tilde{\Lambda}_{\rm GW}$ may allow us to test whether the empirical relation in Eq. \eqref{eq:P/M-Lambda}, derived solely from numerical relativity simulations, actually holds.    
A phase transition from normal nuclear matter to quark matter that can take place inside the NSs around the moment of coalescence might modify the BNS merger process, 
and the tight correlation between $P_{\rm rem}/M_{\rm rem}$ and $\tilde{\Lambda}$ obtained for pure nucleonic stars (as done in this work) may not persist anymore \citep{Bauswein2019}. 
For instance, sharp phase transitions lead to the smaller tidal deformabilities and also induce discontinuities in the relation between tidal deformability and mass \citep{Han2019,Nandi2020}.
Consequently, such phase transitions would lead to a deviation of $\Lambda_{\rm FRB}$-$\Lambda_{\rm GW}$ relation
from that shown in Figure \ref{figure_2}.

\section{Summary and Discussion}
\label{s:discussion}

In this {\it letter}, we investigated the possibility that the separation among the sub-pulses in FRB 181112 could represent the rotation period of an underlying NS, and the extremely short period of about $0.8$ ms could be a strong evidence for a BNS merger.
Base on this picture, we have shown that such a high spin rate inferred from a high time-resolved FRB would offer a unique opportunity to study the nature of the BNS merger remnant, particularly if co-detected with GW. First of all, since the information on the remnant spin period $P_{\rm rem}$ is not yet readily available with the current GW observation, the newly proposed method of detecting it via the high time-resolved FRBs is complementary. 
Moreover, if combined with the remnant NS mass $M_{\rm rem}$ inferred from GW observation, it would place a new constraint on the nuclear matter EOS. Our numerical relativity simulation suggests that the post-merger quantity $P_{\rm rem}/M_{\rm rem}$, or the tidal deformability of the merger remnant, has a tight correlation with the binary tidal deformability parameter $\Tilde{\Lambda}$ of NSs before they merge. Given this empirical relation, a joint FRB-GW observation will establish a new limit on $\Tilde{\Lambda}$. Therefore, if $\Tilde{\Lambda}$ is also well measured by GW data, a comparison between these two will provide further insights into our understanding of nuclear matter and BNS merger process.

Besides the errors related to the simulation described in \S \ref{s:constraints}, there may be additional systematic uncertainties in $M_{\rm rem}$ and $P_{\rm rem}$ that would also propagate to $P_{\rm rem}$-$M_{\rm rem}$ (Figure \ref{figure_1}) and $P_{\rm rem}/M_{\rm rem}$-$\tilde{\Lambda}^{1/5}$ (Figure \ref{figure_2} and Eq. [\ref{eq:P/M-Lambda}]) relations. In this work, we approximated the mass of the remnant NS by the total mass of the pre-merger binary system. Meanwhile, a number of simulations have shown that the post-merger system generally consists of a central core with differential rotation (corresponding to the remnant NS considered here) and an accretion disk that uniformly rotates around it \citep[e.g.][]{Shibata2019}. In this context, the spatial extent of the remnant NS (or the total mass) is not a well-defined concept, but simulations using typical binary masses of $2.5$--$2.7\,M_{\odot}$ suggest that, depending on the mass of the disk, the uncertainty in $M_{\rm rem}$ is up to $0.1$--$0.3\,M_{\odot}$ \citep{Fujibayashi2020}, which translates into a fractional error in $M_{\rm rem}$ of $\lesssim10$\%.

Similarly, $P_{\rm rem}$ may have multiple systematic uncertainties.
First, the spin period is not a gauge-independent quantity in general relativity, and therefore the derived relations could in principle change when choosing different simulation setups. Nevertheless, by comparing the frequency of the dominant quadrupole mode of GW radiation, which is gauge invariant, with the remnant spin frequency we confirm that the effect of gauge is negligible (see Appendix \ref{s:appendix}).
Secondly, since our simulation covers only a limited mass range, we need a more comprehensive study to evaluate the variance of those relations. 
Third, as the rotational profile of the remnant is time-dependent, the minimum spin period may change depending on when one extracts it from a simulation. Finally, the shock-wave heating during the coalescence, which depends on a BNS model, may also affect the rotational profile. 
Since this work is the very first step toward probing BNS merger EOS by means of FRBs, we leave the exploration of these possibilities for future works.

Based on the BNS merger models \citep{Totani2013,Yamasaki2018} and also hinted by the observation of FRB 181112 \citep{Cho2020}, we predict a unique population of non-repeating FRBs having multiple sub-pulses with submillisecond periodicity. The full duration of such FRBs may be determined by the dynamical timescale of ejecta that would hide the radio waves at times of about $0.5$--$1$ ms after the coalescence \citep{Yamasaki2018}. As a result, no subsequent FRB sub-pulse would be observed.

The FRB 121002 \citep{Champion2016} is the first FRB sample that clearly shows double components. However, due to its somewhat large separation between two peaks $2.4\pm0.4$ ms, it cannot be a strong evidence for a BNS merger. Meanwhile, recently discovered double-peaked FRBs 190102 and 190611 \citep{Day2020} could be good candidates for this population. The peak separations for FRBs 190102 and 190611 are about $0.5$ ms and $1$ ms, respectively with small scattering timescales ($0.04$ ms and $0.18$ ms, respectively) and the rotation measure for two sub-pulses in each burst are comparable, sharing many phenomenological similarities to FRB 181112 \citep{Day2020}. Further in-depth modelling of the radio emission signature from merging BNSs \citep[e.g.,][]{Palenzuela2013,Carrasco2020,Most2020,Wada2020} as well as their possible connection to FRBs will be required to see if such models can account for the sub-pulse separations observed.

A population of FRBs that is similar to FRB 181102 will be found by ongoing (ASKAP) and future (e.g., SKA) high-time resolution surveys. Also, there is a fascinating possibility that GWs and radio waves can be accurately observed simultaneously by forecasting the BNS merger with the space-based detector DECi-hertz Interferometer Gravitational wave Observatory (DECIGO, \citealt{Kawamura2006,Sato2017}).  Ultimately, in the era of third-generation detectors, such as the Einstein Telescope \citep[][]{Hild2011} and the Cosmic Explorer \citep[][]{Abbott2017}, post-merger GWs and FRBs similar to FRB 181102 will be detected simultaneously, enabling a direct comparison between the remnant spin periods obtained by FRBs and GWs.

\acknowledgments 
SY gratefully acknowledges the support from the Institute for Cosmic Ray Research during the course of this work. TT was supported by JSPS/MEXT KAKENHI Grant Numbers 18K03692 and 17H06362. KK was supported by 18H01213. The numerical  computation  was  performed  on Cray XC50 at cfca of National Astronomical Observatory of Japan, Oakforest-PACS at Information Technology Center of the University of Tokyo, and on Cray XC40 at Yukawa Institute for Theoretical Physics, Kyoto University.

%






\appendix
\twocolumngrid

\section{Assessment of the Minimum Spin Period Extraction}
\label{s:appendix}

\begin{figure}[t]
\centering
\includegraphics[width=.48\textwidth]{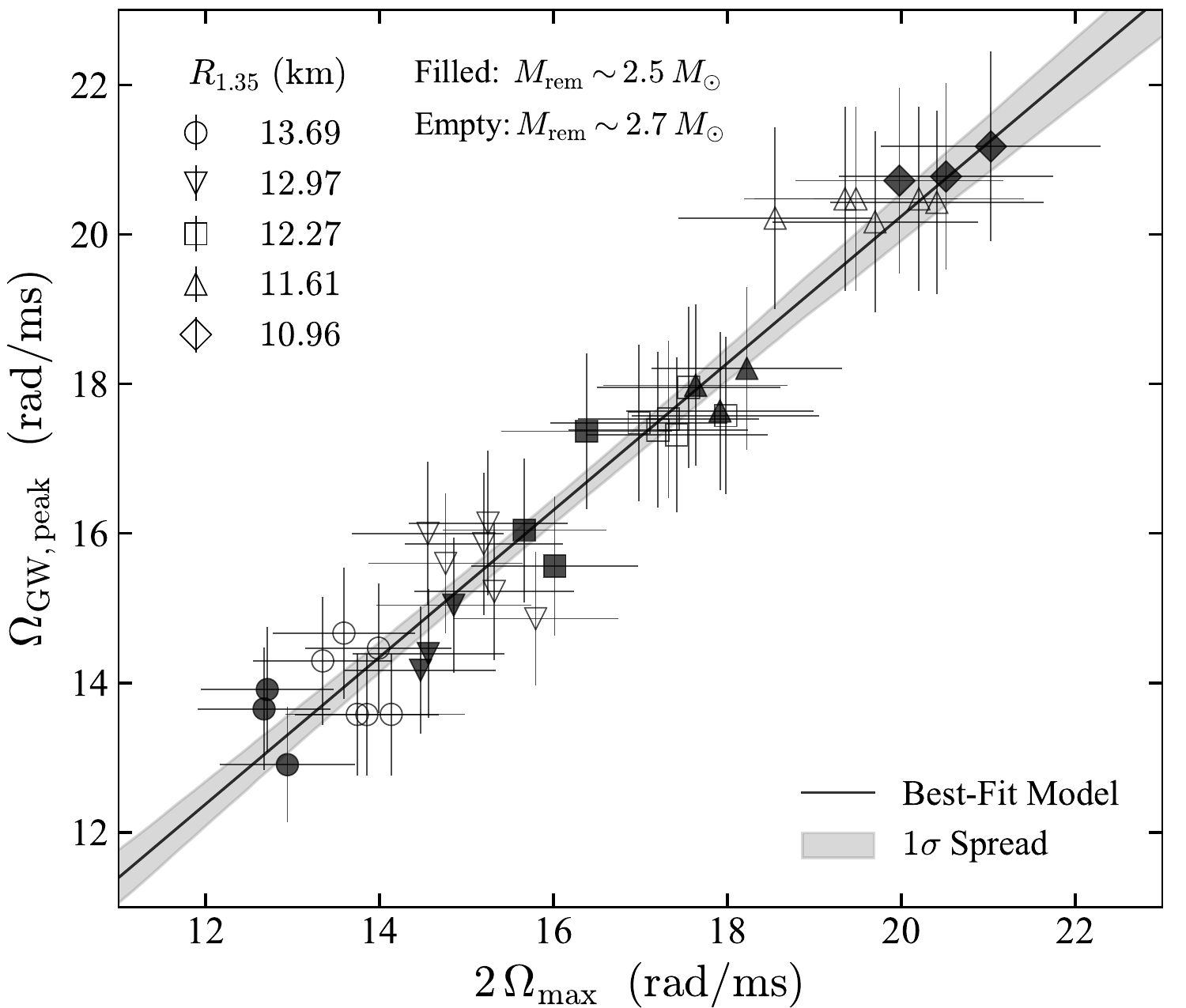}
\caption{The peak in angular frequency space of the post-merger GW spectrum $\Omega_{\rm GW,peak}$ as a function of maximum angular spin frequency of the remnant $\Omega_{\rm max}$. The vertical and horizontal errors are due to the simulation uncertainties, which are set to be $6$\%. The other notations are same as Figures \ref{figure_1} and \ref{figure_2}.
}
\label{figure_3}
\end{figure}

In this work, we extract the minimum spin period of the remnant $P_{\rm rem}=2\pi/\Omega_{\rm max}$ from the simulation by directly examining the peak $\Omega_{\rm max}$ in the angular velocity profile of the remnant after the merger (\S \ref{ss:simulation}). However, one concern is that $P_{\rm rem}$ is not a gauge-invariant quantity and is calculated purely by a Newtonian method. In order to assess the possible gauge dependence of $\Omega_{\rm max}$, we compare it with the gauge-invariant quantity, the peak in angular frequency space of the post-merger GW spectrum $\Omega_{\rm GW,peak}$, 
which is often interpreted as twice the spin frequency of fundamental quadrupole oscillation mode (f-mode) of the remnant NS. Since there is no detailed perturbative calculation of the f-mode frequency for a realistic merged remnant as background (but see e.g.,  \citealt{Kruger2010} for calculations under ideal differentially-rotating background models), it is not yet clear whether the custumary $\Omega_{\rm GW,peak} \approx 2\Omega_{\rm max}$ relation actually holds. Nevertheless, some numerical relativity simulations suggest a relationship between $\Omega_{\rm GW,peak}$ and $2\Omega_{\rm max}$ \citep{Hanauske2017}.
Figure \ref{figure_3} compares $2\Omega_{\rm max}$ calculated from the angular velocity profile with $\Omega_{\rm GW,peak}$ derived from the simulation by a relativistic method using the Weyl scalar \citep{Yamamoto2008}. Clearly, there is an almost linear correlation between them (with a slope of $0.98\pm0.07$ with errors of 1-$\sigma$).
This shows that the influence of the gauge is negligible and $\Omega_{\rm max}$ (or $P_{\rm rem}$) is a sufficiently good physical quantity that represents the minimum spin period of the remnant NS.


\bibliographystyle{apj}



\end{document}